\documentclass{article}

\usepackage{PRIMEarxiv}

\usepackage[utf8]{inputenc} 
\usepackage[T1]{fontenc}    
\usepackage{hyperref}       
\usepackage{url}            
\usepackage{booktabs}       
\usepackage{amsfonts}       
\usepackage{nicefrac}       
\usepackage{microtype}      
\usepackage{lipsum}
\usepackage{graphicx}
\graphicspath{{media/}}     

\title{Reliable Image Transmission in CPS-based Pub/Sub
}

\author{
  Everson Flores \\
  Centro de  Ciências Computacionais \\
  Universidade Federal do Rio Grande\\
  \\
  \texttt{email@email} \\
}

\author{
Everson Flores, Bruna Guterres\textsuperscript{*}, Thomaz Pereira Junior, Paula Barros,\\
Alberto Cabral, Cristiana Lima Dora, Marcelo Malheiros, Marcelo Rita Pias\\
Federal University of Rio Grande (FURG), Brazil\\
\textsuperscript{*}guterres.bruna@furg.br
}

\begin{document}

\maketitle

\begin{abstract}
Developments in communication and automation have driven the expansion of distributed networks, essential for IoT and CPS development in industrial applications requiring reliable image processing and real-time adaptability. Although broadly adopted, there is a literature gap regarding the performance of MQTT  protocol for image sharing and transmission under high-traffic scenarios with intermittent connectivity, restricting its use in critical IoT and CPS applications. In this sense, the present work investigates the reliability of real-time image transmission in IoT and CPS industrial systems that exploit the MQTT-based publish/subscribe communication model. It focuses on scenarios with network interruptions and high data traffic, evaluating the performance of a distributed system through a series of controlled testbed validation experiments. Experimental validation demonstrated that while the MQTT-based system sustains reliable transmission under normal conditions, its recovery capability depends on the failure point, with full restoration occurring when disruptions affect the Orchestrator Node and partial recovery when the Producer Node or Broker are affected. The study also confirmed that the system prevents duplicate errors and adapts well to increasing network demands, reinforcing its suitability for industrial applications that require efficient and resilient data handling. 
\end{abstract}
\section{Introduction}

Improvements in communication and automation technologies have facilitated the expansion of distributed networks and asynchronous communication models, which are fundamental to the development of the Internet of Things (IoT) and Cyber-Physical Systems (CPS) \cite{jiang2021digital, nam2022high} in industrial settings. The fusion of IoT and CPS is particularly relevant for applications requiring reliable image processing and transmission, on which real-time decision-making and system adaptability are crucial. These capabilities rely on lightweight communication protocols for effective data handling. The MQTT (Message Queuing Telemetry Transport) protocol is widely employed in IoT-based networks for its lightweight design and efficiency in low-bandwidth communication, making it well suited for data transfer between resource-limited devices while ensuring scalability \cite{nam2022high}. It also supports reliable real-time communication among various devices \cite{espinosa2015pulga}. However, the demand for real-time data transmission faces significant challenges, particularly in overloaded or unstable network conditions \cite{hwang2017enhanced}, as frequently observed in image-based industrial applications. Despite its widespread adoption, there is a noticeable gap in the literature regarding the performance of the MQTT protocol for image sharing and transmission under high data traffic scenarios and intermittent connectivity which restricts its use in critical IoT and CPS applications \cite{nam2022high, yassein2017internet}.

The present work examines a global publish-subscribe (pub/sub) network architecture leveraging MQTT for image transmission. It evaluates the system performance under diverse traffic loads and disconnection scenarios. This research aims to assess the impact of network disruptions and high data traffic on image throughput and recovery, providing insights into the resilience and robustness of MQTT in complex industrial environments with demanding data requirements. The findings contribute to the enhancement of critical applications, such as environmental monitoring and real-time security networks, on which data reliability and recovery are fundamental \cite{bilal2020transportation, espinosa2015pulga}.

These applications are particularly relevant to industrial water quality monitoring, which is crucial for ensuring compliance with regulatory standards in wastewater management and environmental protection. MQTT-based architectures, with their ability to support efficient and reliable data transmission, may play a key role in industrial wastewater monitoring by enabling real-time data handling, facilitating informed decision-making, minimizing environmental impact, and optimizing operational efficiency. In this sense, the present work presents a case study on image-based microplastic monitoring that addresses a growing environmental concern. Microplastic pollution has captured worldwide attention after reports of massive garbage patches in all ocean gyres \cite{hale2020global} highlighting its widespread and persistent nature \cite{kutralam2023microplastic}. In this sense, effective microplastic monitoring within industrial settings is essential not only to ensure regulatory compliance, but also to reduce industrial contributions to global plastic pollution. The following Research Questions (RQ) are investigated thorugh this case study:

\begin{enumerate}
\item RQ1: What are the key components and configurations for an resilient MQTT-based network architecture that enable reliable image transmission in MQTT-enabled CPS under unstable internet conditions?

\item RQ2: How does the MQTT protocol handle image transmission in scenarios characterized by intermittent connectivity and high network traffic, and what are its limitations for IoT-based industrial applications?

\item RQ3: To what extent can an MQTT-based architecture ensure image integrity, recovery, and synchronization in CPS-driven industrial implementations under varying network conditions?
\end{enumerate}

This paper is organized as follows: Section 2 provides background on MQTT, and IoT-CPS integration. Section 3 describes the modular architecture of the proposed system. Section 4 details the validation testbed, experimental setup, and performance evaluation metrics. Section 5 presents the results and discusses the system’s performance under varying network conditions. Finally, Section 6 concludes the study, summarizing key findings and discussing future research directions.

\section{Related Work}

Cyber-physical systems (CPS) integrate computational and physical processes, enabling software and hardware interaction through sensors, actuators, and communication networks \cite{jiang2021digital}. With advancements in embedded computing and sensor networks, CPS have become essencial in industrial automation, robotics, autonomous vehicles, and smart grids, ensuring precision and efficiency \cite{gomathi2018survey}. The Internet of Things (IoT) further enhances CPS by connecting devices to the internet, supporting remote control, automation, and data-driven optimization within industrial applications \cite{pereira2020challenges, sadeq2019qos}. These IoT-enabled CPS systems rely on lightweight and robust communication protocols such as the Message Queuing Telemetry Transport (MQTT).

MQTT is a pub/sub protocol designed to integrate external data into enterprise systems, suitable for resource-constrained devices operating under limited bandwidth
\cite{alam2016survey}. The pub/sub model enables decoupled communication by organizing messages into topics. Publishers send messages to topics, and subscribers receive them by subscribing, enhancing system scalability and flexibility \cite{maharjan2023benchmarking}. It supports microservices architectures by facilitating asynchronous interactions between components \cite{maharjan2023benchmarking} through technologies such as MQTT, Kafka, and Redis, with MQTT being particularly well-suited for IoT applications due to its efficiency in low-bandwidth networks \cite{nam2022high, maharjan2023benchmarking}.

The MQTT architecture allows indirect interaction through a broker \cite{lima2019performance} and supports one-to-many communication, making it ideal for resource-constrained devices \cite{solanki2017iot}. Its lightweight design is suitable for devices with limited processing and memory, being integrated into IoT and CPS applications due to its asynchronous communication and low power consumption. \cite{espinosa2015pulga, alam2016survey}. The convergence of IoT, MQTT, and CPS drives innovation in sectors such as manufacturing, healthcare, and transportation, enhancing automation and intelligent control \cite{nazir2019reliable, lima2019performance}.

Despite its widespread adoption, data and image transmission in IoT networks still faces challenges sue to packet loss and latency variations, particularly for image-based applications that inherit increased bandwidth needs for efficient data sharing. \cite{silva2021performance} analyzed MQTT, CoAP, and OPC UA, noting MQTT’s higher packet loss under heavy loads due to limited queue support, while CoAP showed increased latency with reduced packet frequency. To address these issues, \cite{hintaw2023robust} proposed the RSS security scheme, which enhances AES with D-AES to improve security and reduce interception losses. RSS also integrates KP-ABE to protect the secret key and enhance access control. Despite these improvements, MQTT continues to face packet loss in high-traffic conditions, reinforcing the need for efficient protocols combined with advanced security mechanisms \cite{hintaw2023robust}.

While existing literature provides valuable insights into MQTT’s performance in general IoT data-sharing scenarios, it offers limited analysis of its suitability for image-based applications, particularly in environments with fluctuating network conditions. In this sense, the present aims to shed some light on the suitability of the MQTT-based pub/sub system for supporting image transmission in CPS applications, assessing its performance under varying network conditions and exploring its potential limitations in high-data-demand scenarios.


\section{System Architecture}
The MQTT-based pub/sub network system is organized in a distributed architecture (Figure 1), in which interconnected nodes perform functions ranging from data acquisition to image restoration, as described in Table \ref{tab:functions_condensed}. Data communication uses the MQTT protocol, explicitly exploring the pub/sub model to support scalability and decoupling between data producers and consumers. This modular design enables the easy addition or replacement of modules without compromising system integrity, providing flexibility in distributed environments.

    \begin{figure}[htb]
    \centering
    \includegraphics[width=0.7\columnwidth]{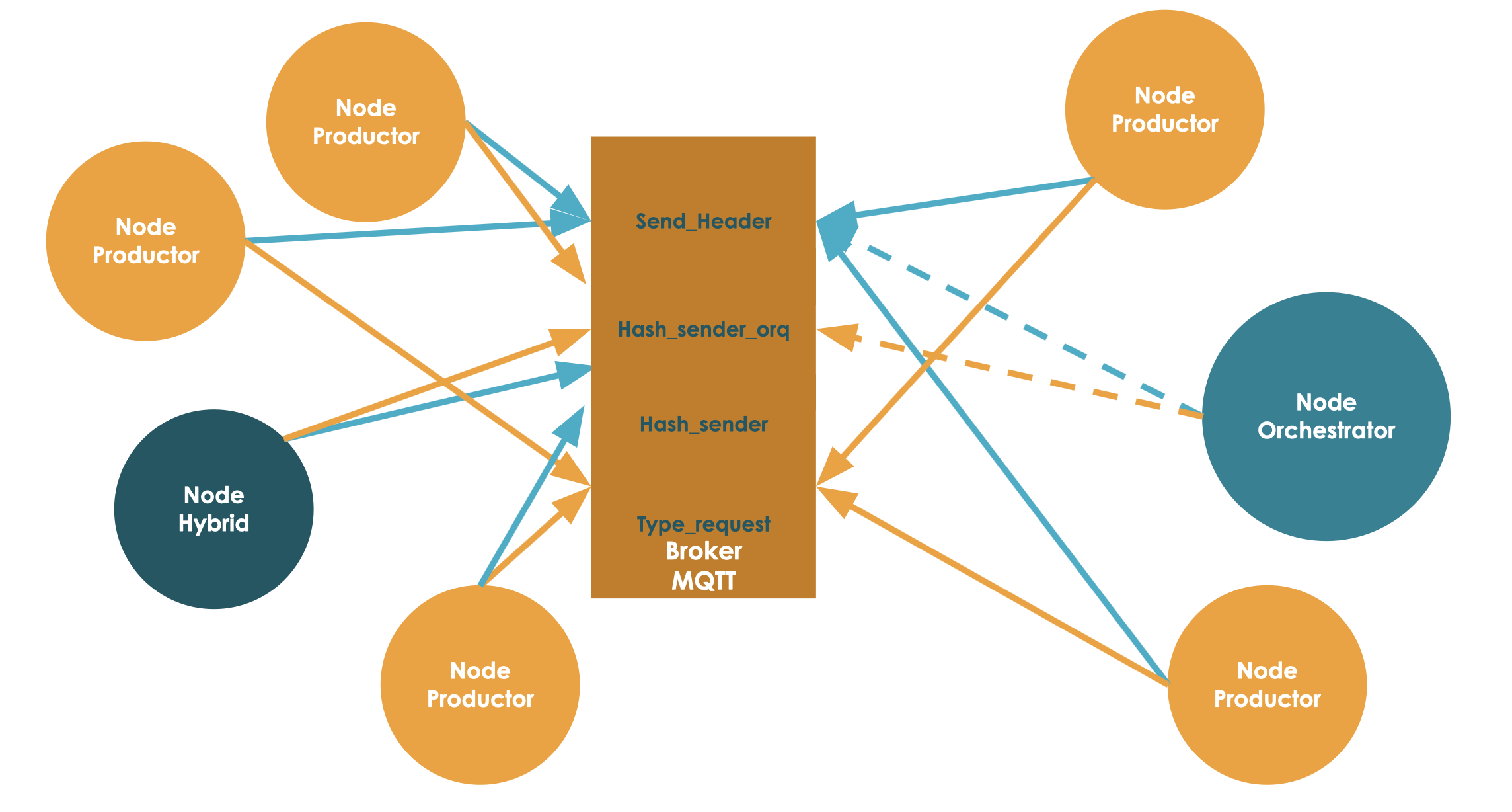}
    \caption{Worldwide Publish-Subscribe Network Architecture Based on the MQTT Protocol.}
    \label{Nodes_arq.}
    \end{figure}


\begin{table}[htb]
  \caption{Functions of Network System Components}
  \label{tab:functions_condensed}
  \centering
  \begin{tabular}{l p{0.6\textwidth}} 
    \toprule
    \textbf{Component} & \textbf{Function} \\
\midrule
Asset & Gathers and distributes data for publication within the network. \\
Producer Node & Acquires and transmits data for processing and storage within the network. \\
Orchestrator Node & Manages network coordination and oversees data storage operations. \\
Hybrid Node & Functions as both a data producer and consumer, supporting dynamic network operations. \\
Image Splitter & Segments images to facilitate transmission over bandwidth-constrained networks. \\
Image Sender & Transmits segmented image fragments via MQTT to a designated network broker. \\
Image Receiver & Receives fragmented images from the broker and processes them accordingly. \\
Image Restore & Reconstructs and stores previously fragmented images. \\
File Status Manager & Oversees data integrity throughout the processing pipeline. \\
Topic Send\_Header & Transmits the initial image header, which includes metadata such as the number of fragments, \textit{hash\_file}, \textit{hash\_sender}, and additional annotations for the message broker. \\
Topic Hash\_sender & Confirms image receipt by returning the \textit{hash\_file} to the \textit{ImageSender} and requests missing fragments when necessary. \\
Hash\_sender\_orq & A dedicated communication topic with the Orchestrator Node, used for transmitting image fragments and handling retransmission requests for missing parts. \\
Storage & Stores processed and reconstructed images. \\

    \bottomrule
  \end{tabular}
\end{table}

\vspace{-10mm}
\section{Validation Methodology}
The validation methodology involves a case study for microplastic monitoring using a CPS that integrates IoT and edge computing for data processing. This system enables fast image sharing. The following sections detail the CPS design, image transmission methodology, and network testbed environment.

\subsection{CPS Design and Component Configuration} 
The proposed CPS integrates components for efficient image capture, processing, and transmission, as described in Table~\ref{tab:cps_configuration}.

\begin{table}[h]
  \caption{CPS Design and Component Configuration}
  \label{tab:cps_configuration}
  \centering
  \begin{tabular}{l p{0.6\textwidth}} 
    \toprule
    \textbf{Component} & \textbf{Description} \\
\midrule
Optical Equipment & The PlanktoScope, an open-access modular flow cytometry platform, was used for microplastic monitoring. The setup includes a Raspberry Pi HQ camera (12.3 MP, Sony IMX477R) with a 25mm M12 lens for improved focus, resolution, and contrast. \\
Computational Environment & Image processing and data management were conducted on computers running Ubuntu OS, supported by Python scripts. \\
Data Communication & The MQTT protocol was implemented to facilitate device communication using the Eclipse Mosquitto broker, known for efficiency in IoT applications. \\
Broker & Hosted on a computer running Ubuntu Server 22.04.1 LTS with an Intel Core i5-6400 processor, optimized for stable and efficient data management. \\
Dataset & Four datasets were used: one large package of 500 MB (445 images) and three smaller packages of 100 MB each, divided into groups of 20 (5 MB), 50 (2 MB), and 100 images (1 MB). \\
    \bottomrule
  \end{tabular}
\end{table}

\subsection{Image Transmission and Reconstruction Process}
The image transmission and reconstruction process follows a point-to-point structure with an MQTT broker connecting the Producer, Orchestrator, and Hybrid Nodes. The Producer Node collects and sends data via MQTT, the Orchestrator receives, validates, and stores images, while the Hybrid Node retrieves stored images to ensure full data access for delayed subscribers, enhancing system resilience. The image transmission process, summarized in Table \ref{tab:image_transmission_moments}, consists of eight key moments designed to improve the system and mitigate packet loss caused by communication failures and connection interruptions.

\begin{table*}[h]
  \caption{Image Transmission Process Step}
  \label{tab:image_transmission_moments}
  \centering
  \begin{tabular}{l p{0.6\textwidth}} 
    \toprule
    \textbf{Step} & \textbf{Description} \\
\midrule
1 - Initial Image Transmission & The \textit{ImageSender} receives the image from an \textit{Asset}, and the \textit{ImageSplitter} partitions it into smaller fragments. A \textit{$hash\_file$} and a metadata header are generated and transmitted via the \textit{SendHeader} topic. The \textit{FileStatusManager} assigns the status "\textit{pending}" to the \textit{$hash\_file$}. If no confirmation is received, the \textit{header} is retransmitted. \\
2 - Header Reception & The \textit{ImageReceiver} extracts the \textit{$hash\_file$} and updates its status to "\textit{working}". The \textit{$hash\_file$} is returned via the \textit{$hash\_sender$} topic, signaling successful reception. \\
\begin{tabular}[c]{@{}l@{}}3 - Confirmation and \\ Transmission  of Parts \end{tabular}& Upon confirmation, the \textit{ImageSender} verifies the \textit{$hash\_file$} status. If accepted, image fragments are transmitted via the \textit{$hash\_sender\_orq$} topic. If rejected, no fragments are sent. \\
\begin{tabular}[c]{@{}l@{}} 4 - Reception of Parts and \\ Attempted Assembly \end{tabular} & The \textit{ImageReceiver} attempts to reconstruct the image using the \textit{ImageRestore} module. Missing fragments trigger Step 5. \\
5 - Request for Missing Parts & The \textit{ImageReceiver} requests missing fragments via the \textit{$hash\_sender$} topic, specifying the \textit{$hash\_file$} and identifiers of the missing parts. \\
6 - Resending Missing Parts & The \textit{ImageSender} retransmits missing fragments via the \textit{$hash\_sender\_orq$} topic. Steps 4 to 6 repeat until successful reconstruction. \\
7 - Successful Assembly Signal & Upon successful reconstruction, the \textit{ImageReceiver} signals completion by transmitting the \textit{$hash\_file$} via the \textit{$hash\_sender$} topic. The \textit{FileStatusManager} updates the status to "\textit{completed}". \\
8 - Finalization and Cleanup & The \textit{ImageSender} verifies the status as "\textit{completed}" and deletes temporary files to conclude the process. \\
    \bottomrule
  \end{tabular}
\end{table*}

The system's implementation is guided by distinct moments, illustrated in Figure \ref{fig:diagrama_moments_2} (A).  

\subsubsection*{\textbf{Sending Images}}  
The Producer or Hybrid Node starts at Step 1 by sending the header to the Orchestrator.  

\subsubsection*{\textbf{On-Demand Requests}}  
\begin{figure}[htbp]
\centering
\includegraphics[width=0.8\columnwidth]{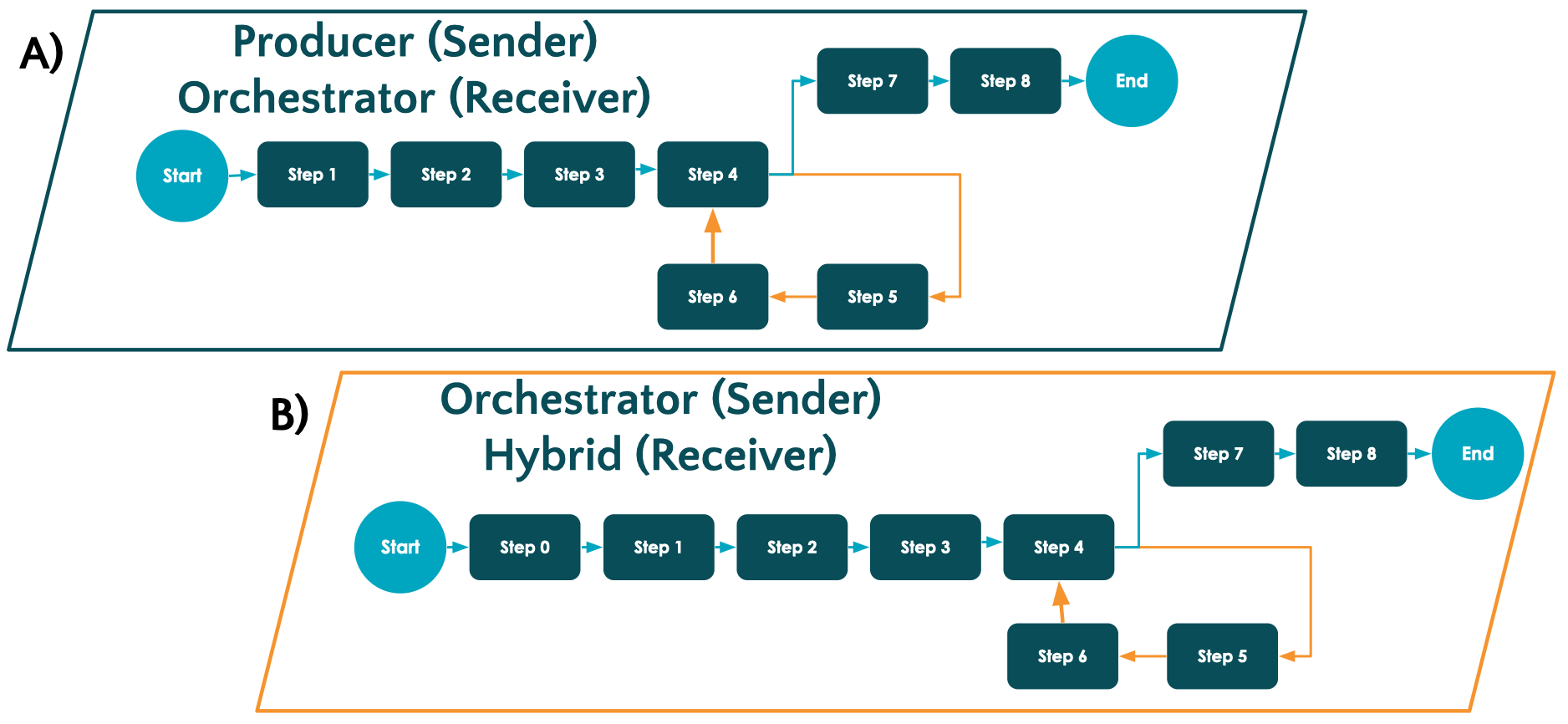}
\caption{Diagram of the steps in the communication between Producer Node and Orchestrator Node, and between Orchestrator Node and Hybrid Node.}
\label{fig:diagrama_moments_2}
\end{figure}

The Hybrid Node first sends a category request (Step 0) via the \textit{$type\_request$} topic, initiating the flow at Step 1 on the Orchestrator side, Figure \ref{fig:diagrama_moments_2} (B).  


\subsection{Network Testbed Environment}
\label{iperf3}

A controlled network testbed is crucial for evaluating CPS performance under varying traffic and interruptions in an IoT network for image transmission. The experiment analyzes transmission behavior and the impact of traffic control software.

\subsection*{Network Architecture and Topology} 

\vspace{-5mm}
\begin{figure}[htb]
\centering
\includegraphics[width=0.7\columnwidth]{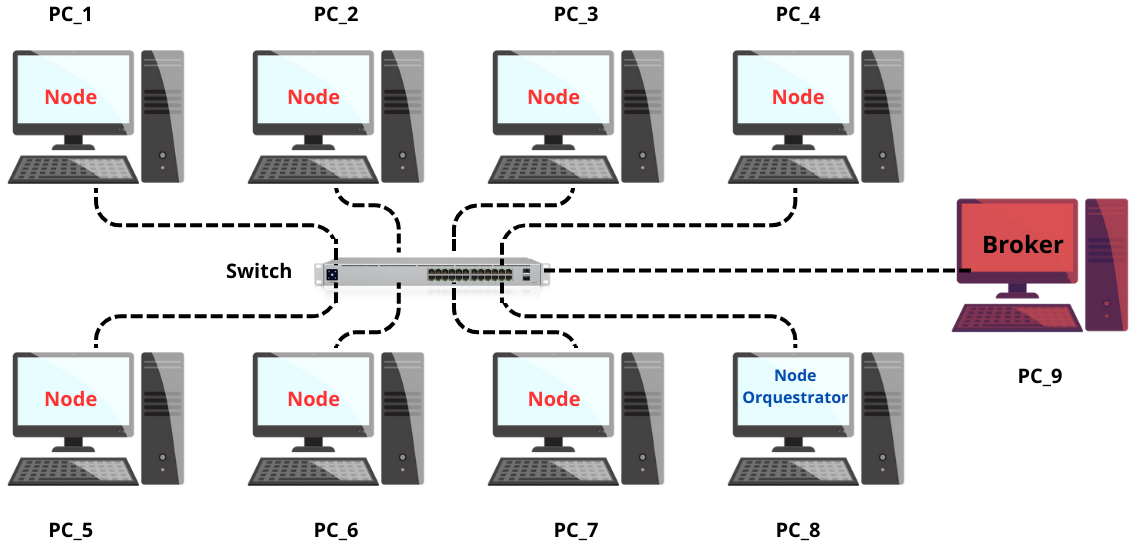}
\caption{Network Testbed Architecture for Image Transmission with Controlled Structure.}
\label{fig:topologia_rede}
\end{figure}

Figure~\ref{fig:topologia_rede} shows the network architecture, consisting of a testbed with nine computers connected to an MQTT broker. Seven act as data producers, sequentially sending images, while one serves as the 'Orchestrator Node,' which receives, stores, and retransmits data. This setup allows for analyzing data flow and MQTT performance in the pub/sub model. The star topology (Figure~\ref{fig:topologia_rede}) connects the devices via a switch, enabling the analysis of image transmission performance over MQTT and interactions between producers and the storage system.

The testbed includes eight computers with Intel® Core™ i3-10100 processors (8 GB RAM), one with an Intel® Core™ i5-6400 processor (6 GB RAM), and a 28-port managed switch. Network performance was evaluated using the \textit{iPerf3} tool, which measures bandwidth, packet loss, retransmissions, and simulates bidirectional traffic to test network behavior under high load.

\subsection*{Performance Parameters and Metrics}
\label{Performance_Parameters_and_Metrics}
The experiments used a 1200 s test duration, up to 8 parallel connections, 131072-byte packets, 500 Mbps bandwidth, and the cubic congestion control algorithm. Additional settings included a 1000 ms rate timer, port 5201, send/receive buffers of 16384/131072 bytes, and pacing control at 62500 kb/s, determining the effective data rate.
 
\subsection*{Data Flow}  
Seven Producer Nodes sequentially send image packets to the Orchestrator Node via an MQTT broker, with random intervals between transmissions. The Orchestrator Node receives, stores, and retransmits the images. Data collection evaluates latency, packet loss, and network processing capacity. The MQTT transmission tests used two packet configurations: one with 445 images totaling 500 MB and another with 170 images totaling 100 MB, composed of 20, 50, and 100 images with sizes of 5, 2, and 1 MB, respectively.

\section{Results and Discussion}
\label{Results and Discussion}
This section presents results and analyzes data transfer performance, focusing on the roles of \textit{Producer}, \textit{Orchestrator}, and \textit{Hybrid} nodes. The evaluation considers file size, transfer frequency, and assesses system effectiveness and resilience using MQTT in a controlled environment.

\subsection{Performance Analysis of Image Transfer}
\label{sec}
 This evaluation presents configurations that simulate potential issues and assess the performance of image transmission using the MQTT protocol under controlled conditions.

\subsubsection*{\textbf{Scenario 1}}  
This scenario involves continuous image packet transmission without internet interruptions or artificial traffic. Table \ref{tab:combined_experiments} shows its structure in eight experiments.

\begin{table}[h]
\caption{Description and Results of Experiments. Results of File Transmission and Image Restoration in Experiments 001 to 015}
\label{tab:combined_experiments}
\centering
\begin{tabular}{ccccccc}
\toprule
\textbf{Exp.} & \textbf{PCs} & \textbf{Pkg (MB)} & \textbf{Imgs} & \textbf{Size (MB)} & \textbf{Sent} & \textbf{Rcvd} \\
\midrule
001 & 1 & 100 & 100 & 1 & 315 & 315 \\
002 & 7 & 100 & 100 & 1 & 315 & 315 \\
003 & 1 & 100 & 50 & 2 & 292 & 292 \\
004 & 7 & 100 & 50 & 2 & 292 & 292 \\
005 & 1 & 100 & 20 & 5 & 235 & 235 \\
006 & 7 & 100 & 20 & 5 & 235 & 235 \\
007 & 1 & 500 & 445 & 5 & 1801 & 1801 \\
008 & 7 & 500 & 445 & 5 & 1801 & 1801 \\
\bottomrule
009 & 1 & 100 & 100 & 1 & 150 & 69 \\
010 & 1 & 100 & 100 & 1 & 150 & 75 \\
011 & 1 & 100 & 100 & 1 & 315 & 315 \\
012 & 7 & 100 & 100 & 1 & 315 & 315 \\
\bottomrule
013 & 2 & 100 & 50 & 2 & 292 & 292 \\
014 & 2 & 100 & 50 & 2 & 292 & 292 \\
015 & 2 & 100 & 50 & 2 & 292 & 292 \\
\bottomrule

\end{tabular}
\end{table}

\vspace{-2mm}

\begin{figure*}[t]  
\centering
\includegraphics[width=\textwidth]{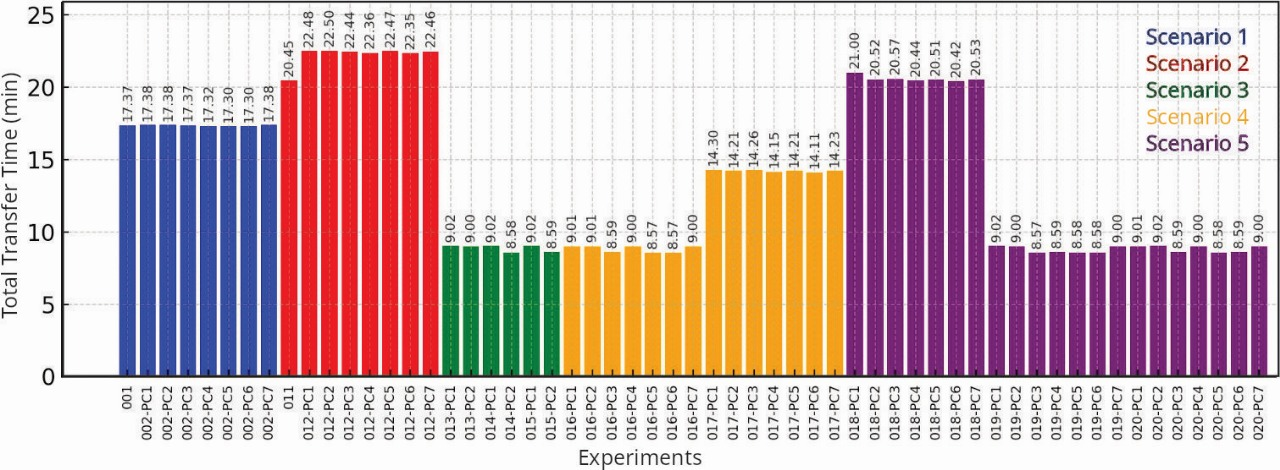}
\caption{Comparative Analysis of Total Transfer Times in a Controlled Network Environment with Image Traffic via MQTT.}
\label{fig:comparative_scenario}
\end{figure*}

In Figure \ref{fig:comparative_scenario}, the transfer time variation was minimal across all groups, indicating consistent results. In groups 001 and 002, times ranged between 17.3 and 17.38 minutes; in groups 003 and 004, between 8.85 and 8.97 minutes; and in group 006, between 3.4 and 3.57 minutes, demonstrating stable transfer conditions. In groups 007 and 008, with larger packets, transfer times were longer (78.13 to 79.18 minutes) but with minimal variation, reinforcing the stability of the experimental environment.

Table \ref{tab:combined_experiments} shows the experiment results. Experiments 001 and 002 achieved 100\% image restoration with 315 files sent and received. In experiments 003 and 004, 292 files were transferred, restoring 50 images each. Experiments 005 and 006 involved 235 files, restoring 20 images. Lastly, experiments 007 and 008 transferred 1801 files, restoring 445 images each. Figure \ref{fig:comparative_scenario} and Table \ref{tab:combined_experiments} confirm consistent performance, minimal transfer time variation, and high restoration rates.

\subsubsection*{\textbf{Scenario 2}} 
The internet connection is interrupted by unplugging the Ethernet cable, allowing evaluation of network disruption impacts and protocol resilience. Table \ref{tab:combined_experiments} details Scenario 2's four experiments.


In experiment 009, the Ethernet cable is disconnected from the Producer Node; in 010, from the broker; in 011, from the Orchestrator Node; and in 012, from the Orchestrator Node during simultaneous image transmission by all seven Producer Nodes. Table \ref{tab:combined_experiments} shows that in experiments 009 and 010, where the Ethernet cable was disconnected from the Producer Node and the broker, 150 files were sent, with 69 and 75 received, resulting in the restoration of 46 and 50 images, respectively. Transmission did not resume after reconnection. In experiments 011 and 012, with disconnection and reconnection at the Orchestrator Node, all 315 files were received, resulting in the complete restoration of 100 images in both cases.

Figure \ref{fig:comparative_scenario} shows consistent performance when the Ethernet cable is reconnected on the Orchestrator Node. In contrast, disconnection on the Producer Node or broker interrupts transmission, resulting in partial image restoration (experiments 009 and 010, Table \ref{tab:combined_experiments}).


\subsubsection*{\textbf{Scenario 3}}

Simulations involve changes in folder names, file names, and image repetition. Table \ref{tab:combined_experiments} outlines Scenario 3's three experiments. In experiment 013, two Producer Nodes send identical images and file names, differing only by package names like "Sample PC1" and "Sample PC2." In experiment 014, the same setup is repeated, with no file duplication observed on the Orchestrator Node. In experiment 015, folders on the Producer Nodes share the same name ("Sample PC1"), yet no folder duplication occurred on the Orchestrator Node. Table \ref{tab:combined_experiments} shows that all files were received in every experiment, ensuring full image restoration without storage duplication. The analysis of Figure \ref{fig:comparative_scenario} shows stable performance with minimal variation in image transfer times across all experiments.

\subsubsection*{\textbf{Scenario 4}} 
The packet transmission experiments were conducted using the \textit{iPerf3} software. The parameters used are detailed in "Performance Parameters and Metrics". Some experiments also involve Ethernet cable disconnection. Table \ref{tab:combined_experiments_016_019} presents the division of Scenario 4 into four distinct experiments.


\begin{table}[h]
\caption{Results of the transmission and restoration of 50 files of 2 MB each in experiments 016 to 019.}
\label{tab:combined_experiments_016_019}
\centering
\begin{tabular}{ccccccc}
\toprule
\textbf{Exp.} & \textbf{PCs} & \textbf{Pkg (MB)} & \textbf{Sent} & \textbf{Rcvd} & \textbf{Downtime (min)} \\
\midrule
016 & 7 & 100 & 292 & 292 & 0 \\
017 & 7 & 100 & 292 & 325 (PC1 \textbf{33}) & 5 \\
018 & 7 & 100 & 292 & 292 & 10 \\
019 & 7 & 100 & 292 & 292 & 0 \\
\bottomrule
\end{tabular}
\end{table}

Table \ref{tab:combined_experiments_016_019} shows the results of experiments 016 to 019, focusing on file transmission, reception, and image restoration. In experiment 016, seven Producer Nodes send images and extra packets to increase traffic, without Ethernet disconnection. In experiment 017, the same setup is used, but the Orchestrator Node’s Ethernet cable is disconnected for five minutes. Experiment 018 repeats this with a ten-minute disconnection, and experiment 019 replicates experiment 018 to confirm results. Figure \ref{fig:comparative_scenario} shows transfer times for experiments 016 to 019. Despite network congestion, times remained consistent with previous experiments, except when the Ethernet cable was disconnected, where delays matched the disconnection duration. In experiment 017, retransmission kept the transfer time within the average.

\subsubsection*{\textbf{Scenario 5}} 
The Hybrid Node comes into play. In this experimental environment, it requests image packets sent by the Producer Node to the Orchestrator Node,. Table \ref{tab:experimentos_020} shows the flow of image requests from the Hybrid Node to the Orchestrator Node.

\begin{table}[htbp]
  \caption{Description of the Image Request Flow Using Hybrid Node to Orchestrator Node, in Experiment 020}
  \label{tab:experimentos_020}
  \centering
  \begin{tabular}{ccccccc}
    \toprule
    \textbf{Hybrid Node} & \textbf{Orch. Node} & \textbf{Files} & \textbf{Files} & \textbf{Images} \\
    \textbf{Requests} & \textbf{Sample} & \textbf{Sent} & \textbf{Received} & \textbf{Restored} \\
    \midrule
    PC 1  & PC 7 & 292 & 292 & 50 \\
    PC 2  & PC 6 & 292 & 292 & 50 \\
    PC 3  & PC 5 & 292 & 292 & 50 \\
    PC 4  & PC 1 & 292 & 292 & 50 \\
    PC 5  & PC 2 & 292 & 292 & 50 \\
    PC 6  & PC 3 & 292 & 292 & 50 \\
    PC 7  & PC 4 & 292 & 292 & 50 \\
    \bottomrule
  \end{tabular}
\end{table}

In experiment 020, all seven computers are configured as Hybrid Node. Each one requests images that have already been sent and stored in the Orchestrator Node,, allowing any Producer Nodes to access samples sent by other Producer Nodes when needed.

The analysis of Figure \ref{fig:comparative_scenario} shows that the application's performance, during image requests by Hybrid Node in the controlled environment, maintains average times comparable to those of image transmissions from Producer Nodes to Orchestrator Node, with minimal variations in receiving times across all Hybrid Nodes in experiment 020.

\section{Conclusions and Future Work}
\label{Conclusions and Future Work}

This work investigated the performance of an MQTT-based architecture for image transmission in IoT and CPS environments under varying network conditions. The evaluation considered scenarios with continuous data flow, intermittent connectivity, and network congestion to determine the resilience of the system to successfully transmit and restore images within a case study that focused on image-based microplastic monitoring for industrial wastewater management.
Under normal operation, image transmission remained stable, demonstrating the protocol’s efficiency in maintaining reliable communication (first scenario). Subsequent scenarios introduced network disconnections, revealing full recovery when the disconnection occurred at the Orchestrator Node, and partial recovery when it occurred at the Producer Node or broker. Experiments with modified file and folder names demonstrated that the system prevents duplication, ensuring accurate data management at the Orchestrator Node. 

Network load and interruption tests using \textit{iPerf3} demonstrated that the system maintained consistent transfer times, with proportional increases corresponding to disconnection durations, highlighting its resilience to more demanding network conditions. Finally, experiments with the Hybrid Node verified the system's capability to efficiently recover previously stored data, with response times similar to those of initial transfers. 
These findings address the research questions by showing that the MQTT-based architecture can support reliable image transmission (RQ1), effectively handle high network traffic and intermittent connectivity (RQ2), and ensure image integrity and recovery in industrial CPS applications (RQ3).

The results suggest that the proposed architecture is well suited for image-based real-time monitoring applications, including industrial water quality assessment, where image transmission reliability is critical. While the system demonstrated effective data recovery and transmission stability. The main conclusions include: (1) The MQTT architecture performed effectively under controlled interruption conditions, successfully recovering transmission in most cases, especially when the failure occurred in the Orchestrator Node; (2) MQTT's performance under high load remained consistently good, restoring data integrity quickly, though with some increase in latency; and (3) The system's ability to reconstruct fragmented images and transmit them efficiently highlights the robustness of the architecture, making it suitable for real-time monitoring.

Further research could explore optimizations to enhance its adaptability in more complex industrial deployments, enhance notification systems to alert users about new data availability and integrate artificial intelligence into the framework for predictive data analysis.

\section*{Acknowledgment}
This work is part of the ASTRAL (All Atlantic Ocean Sustainable, Profitable and Resilient Aquaculture) project. This project has received funding from the European Union’s Horizon 2020 research and innovation programme under grant agreement Nº 863034.

\bibliographystyle{IEEEtran}
\bibliography{sample-base}

\end{document}